\documentclass{article}
\usepackage{spconf,amsmath,graphicx,bm,url,fancyvrb,booktabs}

\newcommand{\weight}[1]{w_{\text{#1}}}
\newcommand{\xx}[1]{\bm{x}^{(#1)}}
\newcommand{\y}[2]{\bm{y}_\text{#1}^{(#2)}}
\newcommand{\haty}[2]{\bm{\hat y}_\text{#1}^{(#2)}}

\usepackage{ifthen}
\newcommand{\xdetail}[1]{\{\bm{x}^{(#1)}[n]\}_{n=0}^{N^{(#1)}-1}}
\newcommand{\Lossnoarg}[1]{\mathcal{L}_\text{#1}}
\newcommand{\ydetailelement}[2]{\ifthenelse{\equal{#1}{F0}}{y}{\bm{y}}_{\text{#1}}^{(#2)}}
\newcommand{\ydetailelementscalar}[2]{y_{\text{#1}}^{(#2)}}
\newcommand{\hatydetailelementscalar}[2]{\hat{y}_{\text{#1}}^{(#2)}}

\newcommand{\hatydetailelement}[2]{\ifthenelse{\equal{#1}{F0}}{\hat{y}}{\bm{\hat{y}}}_{\text{#1}}^{(#2)}}
\newcommand{\ydetail}[2]{\{\ydetailelement{#1}{#2}[t]\}_{t=0}^{T-1}}
\newcommand{\hatydetail}[2]{\{\hatydetailelement{#1}{#2}[t]\}_{t=0}^{T-1}}
\newcommand{\tildex}[1]{\bm{\tilde{x}}^{(#1)}}
\newcommand{\VoicedIndexSet}[1]{\mathcal{V}^{(#1)}}

\title{DNN-based ensemble singing voice synthesis \protect\\with interactions between singers}
\name{Hiroaki Hyodo$^1$, Shinnosuke Takamichi$^{2, 1}$, Tomohiko Nakamura$^3$, Junya Koguchi$^4$, Hiroshi Saruwatari$^1$}
\address{
  $^1$The University of Tokyo,  Japan\\
  $^2$Keio University,  Japan\\ 
  $^3$National Institute of Advanced Industrial Science and Technology (AIST), Japan\\
  $^4$Meiji University, Japan
}

\begin{document}
%
\maketitle
\begin{abstract}
    We propose a singing voice synthesis (SVS) method for a more unified ensemble singing voice by modeling interactions between singers.
    Most existing SVS methods aim to synthesize a solo voice, and do not consider interactions between singers, i.e., adjusting one's own voice to the others' voices. Since the production of ensemble voices from solo singing voices ignores the interactions, it can degrade the unity of the vocal ensemble.
    Therefore, we propose a SVS that reproduces the interactions. It is based on an architecture that uses musical scores of multiple voice parts, and loss functions that simulate the interactions' effect to acoustic features. Experimental results show that our methods improve the unity of the vocal ensemble.
    
\end{abstract}

\begin{keywords}
singing voice synthesis, deep learning, vocal ensemble, a cappella
\end{keywords}
\section{Introduction}
\label{sec:intro}
\begin{figure}[t]
        \centering
        \includegraphics[width=0.85\linewidth]{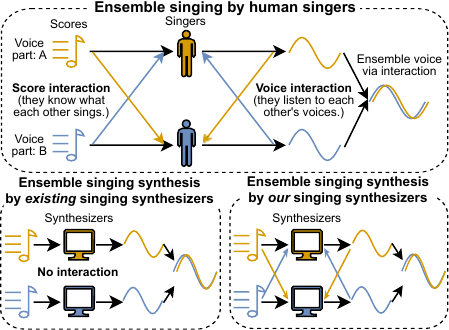}
        \caption{Concept of proposed ensemble SVS approach explicitly modeling interactions between singers. 
        Human singers adjust their voices by listening to the others' voice.
        Conventional SVS methods separately synthesize singing voices for each voice part. In contrast, proposed method produces ensemble singing voices using information of other voice parts as human singers interact.
        }
        \label{fig:concept}
    \end{figure}

    Singing is familiar to many people as a means of communication and artistic expression.
    A group of multiple singers singing simultaneously is referred to as \emph{vocal ensemble}.
    A distinct feature of ensemble singing from solo singing is interactions between singers. The singers comprising vocal ensemble adjust their own voices while listening to the others' voices~\cite{Devaney2012ISMIR,dai2017analysis, Cuesta2018ICMPC,Dai2019JASA,Weiss2019ISMIR,Rosenzweig2020TISMIR_1,Kambara2022}, as shown in the upper part of Fig.~\ref{fig:concept}.
    This adjustment harmonizes the voices and establishes a sense of unity in the ensemble singing.
    Thus, the interactions between singers are necessary to realize a unified ensemble singing.

    Singing voice synthesis (SVS) aims at producing singing voice by computers. Its performance has greatly improved owing to deep neural networks (DNNs)~\cite{hono2021sinsy,liu2022diffsinger,CeVIO,NEUTRINO}.
    %
    Despite the success of the DNN-based SVS methods, most existing methods focus solely on synthesizing a solo voice. The interactions between singers have been overlooked and have not been taken into account, which may result in a lack of unity in ensemble singing voice (as shown in the lower left of Fig.~\ref{fig:concept}).
    By considering the interactions we can enhance the unity of the synthesized ensemble singing voices and elucidate of how humans perceive unity within ensemble voices.
    
    In this study, we propose an ensemble singing voice synthesis method that reproduces the interactions among human singers, as shown in the lower right of Fig.~\ref{fig:concept}.
    As in most of the vocal ensemble studies \cite{Cuesta2020ISMIR,Gover2020ISMIR,Petermann2020ISMIR,Sarkar2021Interspeech,nakamura2023jacappella}, our focus is on interactions within vocal ensembles featuring one singer per voice part, where each singer may perform a different melody. This problem is the generalized version of singing voice synthesis where singers perform the same melody (e.g., unison~\cite{chandna2022deep}).
    We attempt to realize interactions at the level of musical score and acoustic features by multi-track architecture and loss functions, and conduct experimental evaluations to assess the influence of them on synthesized singing voices.

\section{Related works} \label{Sec:関連研究}
    \subsection{Interaction among singers}
        Practical guidebooks for vocal ensemble beginners \cite{asshiacappella, asshiacappellaadvance} provide some instructions of the interactions between singers.
        For example, ``you should sing in a way that your own voice blends in and disappears when mixed with the voice of another singer, keeping the musical score of the other singers in mind,''~\cite{asshiacappella} and ``you should balance the volume of all singers.''~\cite{asshiacappellaadvance}.
        
        The interactions influence acoustic features of singing voices~\cite{dai2017analysis, Kambara2022}.
        For example, Dai et al.~\cite{dai2017analysis} and Kambara~\cite{Kambara2022} have showed that listening to the other singers' voices changes the fundamental frequency ($F_0$) and format frequencies of the singer.
        These facts are helpful for ensemble singing voice synthesis.
        
    \subsection{Evaluation metrics of unity of vocal ensemble}
    \label{subsec:eval_metrics}
        The interactions between singers enhance the unity in the ensemble singing.
        However, there is no standard evaluation metrics on the unity of synthesized singing voices.
        Kambara~\cite{Kambara2022} has proposed six criteria (oneness, pitch, breathing, resonance, voice harmony, and vowel) to evaluate the unity of human ensemble singing voices. The author has elucidated the correlations between listeners' preference according to each criterion and acoustic features.
        Thus we use these criteria in the subjective evaluation in this paper.

    \subsection{Singing voice synthesis}
        Recent SVS methods employ DNNs to learn the relationship between musical score features and singing voices~\cite{hono2021sinsy, liu2022diffsinger}. 
        In this paper, we adopt the commonly used tandem, vocoder-based framework~\cite{hono2021sinsy} as the backbone model. This framework consists of a musical score parser, a time-lag model that predicts onset timing of each note in a musical score, a duration model that predicts  phoneme durations of each note, an acoustic model, and a vocoder.
        Although end-to-end frameworks~\cite{zhang2022visinger} and the use of self-supervised learning features~\cite{jayashankar2023self} are more state-of-the-art technologies, we do not adopt these in this paper. This is because these technologies are black-boxed, making it challenging to associate them with the insights from analysis of vocal ensemble discussed so far. Therefore, we use an explicit inference structure and parametric features.

    \subsection{Chorus synthesis based on signal processing}
        Some previous studies artificially create ensemble singing voices by modifying a solo singing voice rather than synthesizing from scores.
        For example, Petermann et al.~\cite{Petermann2020ISMIR} has proposed a method that produces unison voices by adding up solely synthesized or converted singing voices.
        Thus the interactions between singers are not modeled.
        
\section{Method}
    We propose an ensemble singing voices synthesis method that explicitly models the interactions between singers at the levels of musical scores and acoustic features.
    The proposed network architecture uses not only the musical scores of the target voice part but also those of other voice parts to predict acoustic features (Section~\ref{subsec:arch}). This corresponds to the score-level interaction that each singer keeps the musical contents of the other voice parts in mind.
    As acoustic features we use mel-cepstral coefficients (MGC), continuous log $F_0$ (LF0), band aperiodicity (BAP), and voiced/unvoiced flags (V/UV), each type of which is processed by a separate DNN.
    We also propose loss functions across voice parts (Section~\ref{subsec:loss}). It corresponds to the acoustic-feature-level interaction that each singer adjusts their voice to match the voices of the other singers.
    For simplicity, we assume that the number of voice parts is two (Voice part A and Voice part B) in the following.
    We note that our proposed method can be extended to the case of more than two voice parts as we will discuss in Section~\ref{subsec:3part}.
    \subsection{Data preprocessing}
    \label{前処理：データ分割}
        In the training of DNN-based solo singing synthesizers, loading entire song data into a mini-batch at once is not practical due to GPU memory limitations. 
        To avoid this problem, we often divide features before and after periods of silence and treat the divided features as individual samples. 
        This method is valid when model of each voice part is independently trained. However, when the models for multiple voice parts are jointly trained, the segmentation must be synchronized across the voice parts. Thus, we divide the features at the times when all the voice parts are silent to obtain synchronized features.

        Let $i \in \{\text{A},\text{B}\}$ be the voice part index and $\xx{\text{i}}:=\xdetail{\text{i}}$ be the sequence of score features for voice part $i$.
        Here $N^{(i)}$ represents the length of the sequence for voice part $i$.
        The sequences of $F_0$ and mel-cepstrum for voice part $i$ are denoted as $\y{F0}{i}:=\ydetail{F0}{i}$ and $\y{mgc}{i}:=\ydetail{mgc}{i}$, respectively, where $T$ is the sequence length of $F_0$ and mel-cepstrum.
        Since $\xx{\text{A}}$ and $\xx{\text{B}}$ may differ in sequence length, we match their lengths using a padding method. The padded sequences are denoted as $\tildex{\text{A}}$ and $\tildex{\text{B}}$, respectively.

        One simple padding method to handle multiple sequences is to extend the conventional padding method. It pads the end of shorter sequences in the same way as general mini-batch training.
        We call this method \textit{post-padding}.
        Despite its simplicity, it ignores the synchrony across voice parts.
        In vocal ensemble songs, some notes are sung synchronously across the singers.
        For example, when the notes appear simultaneously across voice parts in the musical scores, they should be sung with the same onset times.
        Thus, when using the post-padding method, the score features of such notes may not be present in the same time slots across voice parts.

\begin{figure}[t]
            \centering
            \includegraphics[width=1.00\linewidth]{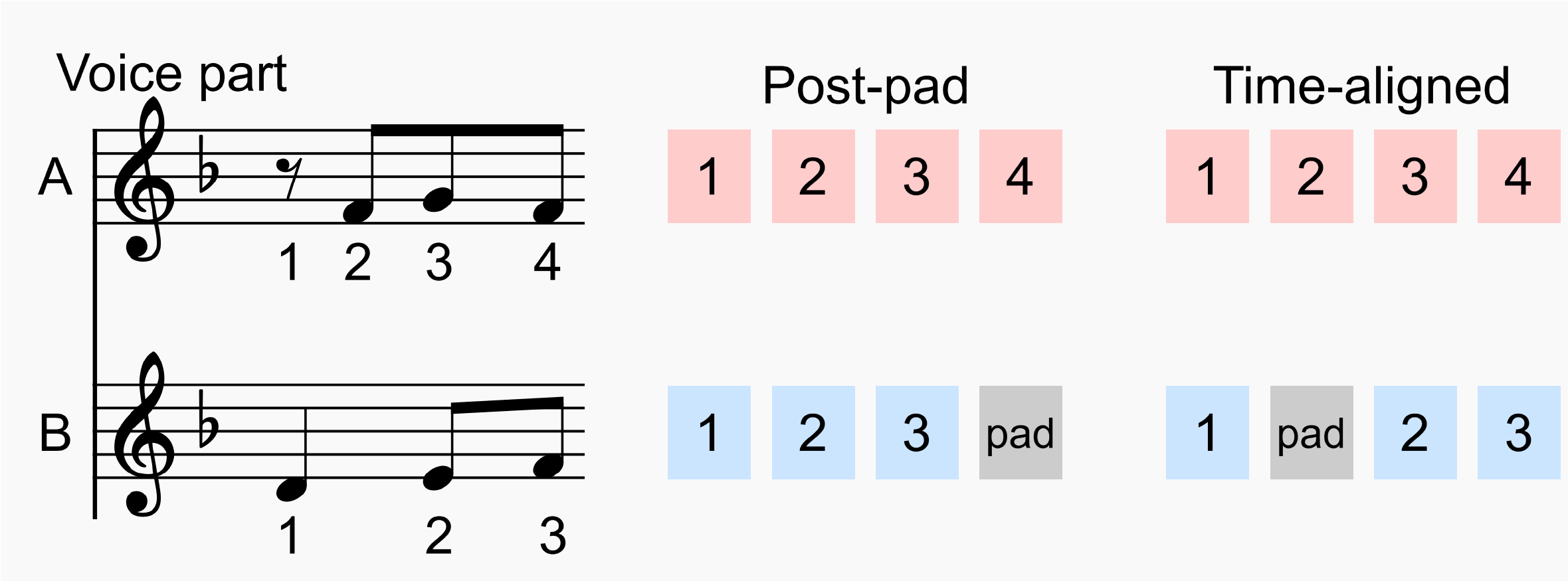}
            \vspace{-5pt}
            \caption{Two padding methods of score features. Numbers indicate note indices.}
            \label{fig:padding}
        \end{figure}

        To capture this synchrony, we propose a synchrony-aware padding method, \textit{time-aligned padding} (see Fig.~\ref{fig:padding}).
        It pads $\xx{\text{i}}$ so that the notes whose onset times are the same between the voice parts are present in the same time slots of $\tildex{\text{i}}$.
        Owing to this characteristic, the proposed padding method captures the synchrony across voice parts well.

    \subsection{Architecture}
    \label{subsec:arch}
\begin{figure}[t]
            \centering
            \includegraphics[width=1.0 \linewidth]{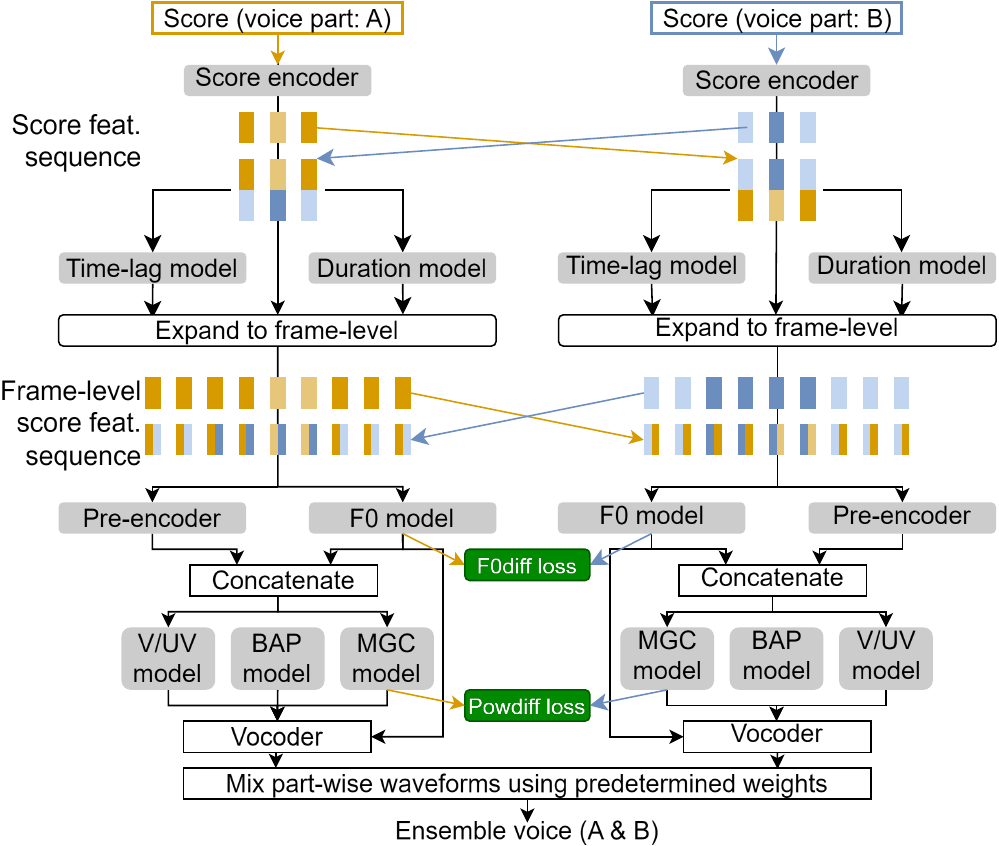}
            \caption{Network architecture and loss functions of proposed method. For brevity, singer-embeddings and conventional loss functions are omitted.}
            \label{fig:architecture}
        \end{figure}

        Fig.~\ref{fig:architecture} shows the architecture of the proposed method.
        It is composed of multi-track parallel structures, each of which accounts for one voice part.
        Each structure has time-lag, duration, and acoustic models, which interact with those of the other voice part.
        These models may take singer-embeddings as additional inputs to handle multiple singers.
        The other modules of the structure, a score encoder and a vocoder, are agnostic to the voice parts.
        

        For the time-lag and duration models, we concatenate $\tildex{\text{A}}$ and $\tildex{\text{B}}$ along the feature axis, and input them to predict time-lags and durations in units of frames. Then the predicted ones are used to expand the score features along the time axis. The expanded ones are added between voice parts and fed into the acoustic model.

        For the acoustic model, we first input the sequence of score features expanded at the frame level to the pre-encoder, and the log $F_0$ model that predicts the frame-level sequence of log $F_0$s. Next, the frame-level score sequence and the log $F_0$ sequence are inputted to the models to predict mel-cepstrum, band aperiodicity and voiced/unvoiced flags.
        In this paper, we use two types of acoustic models: a long short-term memory (LSTM)~\cite{hochreiter1997long}-based model and a diffusion-based model.

        \smallskip
        \textbf{LSTM-based acoustic model.} 
            The LSTM-based acoustic model~\cite{hono2021sinsy} consists of pre-encoder, log $F_0$ model, mel-cepstrum model, band aperiodicity model, and voiced/unvoiced flag model.
            The pre-encoder consists of 3 LSTM layers and one linear layer.
            The log $F_0$ model consists of feed-forward layers, convolutional layers, LSTM layers, and a duration-informed Tacotron with residual F0 prediction~\cite{okamoto2019tacotron}.
            Mel-cepstrum model, band aperiodicity model, and voiced/unvoiced flag model consist of a linear layer, a convolutional layer, LSTM layers, and a linear layer.


        
        \textbf{Diffusion-based acoustic model.} 
            The diffusion-based acoustic model consists of log $F_0$ model, mel-cepstrum model, band aperiodicity model, and voiced/unvoiced flag model.
            Mel-cepstrum model and band aperiodicity model are the diffusion models~\cite{liu2022diffsinger}, which has one conditioning features encoder composed of a linear layer, a convolutional layer, LSTM layers, and a linear layer.
            The log $F_0$ and voiced/unvoiced models are the same as those of the LSTM-based acoustic model.

        \smallskip
 
    \subsection{Loss functions} 
    \label{subsec:loss}
        For the time-lag and duration models, we use the mean squared error (MSE) loss between the ground-truth and predicted values for each voice part~\cite{hono2021sinsy}.
        In the LSTM-based acoustic model, we use the mean absolute error (MAE) loss between ground-truth and predicted features.
        In the diffusion-based acoustic model, we use the MAE loss between noises added to the ground-truth and those predicted by the model~\cite{liu2022diffsinger}.
        Among the loss functions of the above acoustic models, the one related to the log $F_0$ model is denoted as $\Lossnoarg{F0}$, and the one related to the mel-cepstrum model is denoted as $\Lossnoarg{mgc}$.

        To induce the interaction at the acoustic feature level, we propose two loss functions based on acoustic feature differences across voice parts.
        Let $\haty{F0}{i}:=\hatydetail{F0}{i}$ and $\haty{mgc}{i}:=\hatydetail{mgc}{i}$ be the predicted $F_0$ and mel-cepstrum sequences for voice part $i$, respectively.

        \smallskip
        \textbf{F0 difference loss.} The difference in $F_0$ between singers changes by the interactions (as described in Section~\ref{Sec:関連研究}). We define the MAE loss between the ground-truth and predicted values for the difference between two $F_0$ sequences at voiced frames:
            \begin{align}
                \Lossnoarg{F0diff} =& \sum_{t\in\VoicedIndexSet{\text{A}}\cap\VoicedIndexSet{\text{B}}}
                \left|
                    \Delta_{\text{F0}}[t] - \hat{\Delta}_{\text{F0}}[t]
                \right|
                , \label{eq:f0diff}
                \\
                \Delta_{\text{F0}}[t]:=&\ydetailelement{F0}{\text{A}}[t] - \ydetailelement{F0}{\text{B}}[t]
                ,
                \\
                \hat{\Delta}_{\text{F0}}[t]:=&\hatydetailelement{F0}{\text{A}}[t] - \hatydetailelement{F0}{\text{B}}[t]
                ,
            \end{align}
        where $\VoicedIndexSet{i}$ denotes the set of voiced frame indices in voice part $i$.
        This loss function induces the coherence in $F_0$ across voice parts.
        More specifically, the $F_0$ curves change synchronously over time, which leads to better harmonization of the synthesized singing voices.

        \textbf{Power difference loss.} Similarly, the interactions between singers also affect the power of the singing voices (see Section~\ref{Sec:関連研究}).
        The loss to represent this interaction is defined as
            \begin{align}
                \Lossnoarg{Powdiff} =&
                \sum_{t\in\VoicedIndexSet{\text{A}}\cap\VoicedIndexSet{\text{B}}}
                \left|
                    \Delta_{\text{mgc}}[t,0] - \hat{\Delta}_{\text{mgc}}[t,0]
                \right|
                , \label{eq:powdiff}
                \\
                \Delta_{\text{mgc}}[t,0]:=&\ydetailelementscalar{mgc}{\text{A}}[t,0] - \ydetailelementscalar{mgc}{\text{B}}[t,0]
                ,
                \\
                \hat{\Delta}_{\text{mgc}}[t,0]:=&\hatydetailelementscalar{mgc}{\text{A}}[t,0] - \hatydetailelementscalar{mgc}{\text{B}}[t,0]
                ,
            \end{align}
        where $\ydetailelementscalar{mgc}{i}[t,0]$ and $\ydetailelementscalar{mgc}{i}[t,0]$ represent the ground-truth and predicted zeroth MGCs of voice part $i$ at time $t$, respectively.
        This loss biases to the synchronous power changes over time, which suppresses the unintentional prominence of the singing voice of a specific voice part in the ensemble singing.
        \smallskip

        To sum up, the resultant loss $\mathcal{L}$ is
        \begin{equation}
        \mathcal{L}=\Lossnoarg{F0}+\Lossnoarg{mgc}+\weight{F0diff}\Lossnoarg{F0diff}+\weight{Powdiff}\Lossnoarg{Powdiff}
            ,
        \end{equation}
        where $\weight{F0diff}$ and $\weight{Powdiff}$ represent the weights for $\Lossnoarg{F0diff}$ and $\Lossnoarg{Powdiff}$, respectively.

    \subsection{Inference} 
    \label{subsec:gen_acappella}
        The waveforms of individual voice parts are sequentially synthesized (see Fig.~\ref{fig:architecture}).
        First, the score features of the target voice part are concatenated with those of the other voice parts along the feature axis. 
        The concatenated score features are fed into the time-lag and duration models.
        Then, on the basis of their outputs, the score features are expanded to the frame-level score features.
        The frame-level score features of each voice part are added to those of the other voice part and are fed into the acoustic model to predict the acoustic features.
        The obtained acoustic features of each voice part are converted into the waveform of that voice part by the vocoder. 
        Finally, all the waveforms are summed up to form the ensemble singing voice waveform using predetermined weights.

    \subsection{Extensibility to more than two voice parts} \label{subsec:3part}
        We have thus far discussed the case of two voice parts, but some vocal ensemble songs have more than two voice parts (e.g., see \cite{nakamura2023jacappella, jeon2023medleyvox}).
        In this section, we discuss extensions of the proposed method to vocal ensemble songs with $N>2$ voice parts.
        (Note that in the experimental evaluations we only deal with two voice parts.)

        One straightforward approach to handle more than two voice parts is to consider all the combinations of $N$ voice parts, i.e., use $N(N-1)$ F0 difference loss and $N(N-1)$ power difference loss.
        It can handle the interactions between all singers and offer the consistent training and inference pipelines, while its computational cost increases combinatorially.

        Another approach is to combine the proposed two-voice-part method with random selection of voice parts.
        During training, it generates a mini-batch from a lead vocal voice part and randomly chosen one of the other $N-1$ voice parts. Thus, we can use the same training method as the case of two voice parts.
        It can avoid the combinatorial increase in computational cost, while the interactions between the singers other than the lead vocal singer are not directly handled.

\section{Experimental evaluation}
    \subsection{Experimental conditions} \label{sec:exp_cond}
        As datasets we used the jaCappella corpus~\cite{nakamura2023jacappella} for the models to learn ensemble singing voices, and Namine Ritsu Japanese singing voice database Ver.~2~\cite{canon_ritsu} to enlarge training data size.
        The jaCappella corpus comprises 35 Japanese a cappella vocal ensemble songs. We annotated phoneme alignments\footnote{We will publish the alignment results after acceptance.} to the lead vocal and soprano voice part of 15 jaCappella songs (23 minutes) and used three songs as the test set and the remaining 12 songs as the training and validation sets.
        The singer IDs of the annotated lead vocal and soprano are Vo1 and S1, respectively.
        Namine Ritsu Japanese singing voice database Ver.~2 comprises annotated Japanese singing voices of 110 songs with a single female singer. 
        We used 101 songs (371 minutes) as the training and validation sets.
        
        For mixing part-wise voices into an ensemble singing voice, we added the lead vocal and soprano waveforms with average power ratio of $1:2/3$\footnote{We ensured that both parts could be adequately perceived.}.
        
        

        All singing voices were sampled at $48$~kHz.
        We used WORLD~\cite{morise2016world} (D4C edition~\cite{morise2016d4c}) to extract the acoustic features: mel-cepstral coefficients ($60$ dim.), continuous log $F_0$ ($1$ dim.), band aperiodicity ($5$ dim.), and voiced/unvoiced flag ($1$ dim.).
        For the WORLD analysis, we used HARVEST~\cite{morise2017harvest} with a frame shift of 5 ms.
        The dimension of singer embeddings was $256$.
        As the score encoder, we used NEUTRINO~\cite{NEUTRINO} musical score parser.
        Our implementation followed NNSVS~\cite{yamamoto2023nnsvs}\footnote{We will publish the code as the NNSVS recipe after acceptance.}. 
        The network architecture and hyperparameters used in the LSTM-based acoustic model and the diffusion-based model are the same as \emph{MultistreamSeparateF0ParametricModel}\footnote{\url{https://github.com/nnsvs/nnsvs/blob/master/recipes/amaboshi_cipher_utagoe_db/dev-48k-world/config.yaml}} and \emph{NPSSMDNMultistreamParametricModel}\footnote{\url{https://github.com/nnsvs/nnsvs/blob/master/recipes/namine_ritsu_utagoe_db/dev-48k-world/config.yaml}} in the NNSVS library~\cite{yamamoto2023nnsvs}, respectively.
        The diffusion-based model is trained with 100 sampling steps.
        We used Adam ($\alpha=0.0001$, $\beta_1 = 0.9$, $\beta_2 = 0.999$)~\cite{kingma2014adam} for optimization.

    \subsection{Acoustic model: LSTM vs. diffusion} \label{subsec:音響モデルの比較}
        Before examining the effect of the proposed method, we compared the LSTM- and diffusion-based acoustic models through a solo SVS task to ensure better singing voice synthesis quality.
        We used the same data as described in Section~\ref{sec:exp_cond}.
        The synthesized singing voice waveforms were evaluated using a five-point mean opinion score (MOS) test on the naturalness.
        30 listeners participated, and each listener evaluated 12 phrases of singing voices. 

        \begin{table}[t]
    \centering
    \caption{MOS of part-wise voices with $95$\% confidence intervals. LSTM is superior with $p < 0.05$}
    \vspace{3mm}
    \label{table:LSTMvsDiffusionMOS}
    \footnotesize
    \begin{tabular}{c|c} \toprule
        Method & MOS $(\uparrow)$ \\ 
        \midrule
        LSTM & $\textbf{2.82}\pm0.15$ \\ 
        Diffusion & $2.56\pm0.15$ \\ \bottomrule
    \end{tabular}
 \end{table}

        Table~\ref{table:LSTMvsDiffusionMOS} shows the MOS scores of the two acoustic models. The LSTM-based acoustic model had higher MOS scores on average with statistical significance. Thus, we used the LSTM-based model in the following experiments.

    \subsection{Padding: post-pad vs. time-aligned}
        We compared the post-pad and time-aligned padding methods in terms of prediction errors of the time-lag and duration models.
        These methods used the score features concatenated along the voice parts.
        As a baseline, we also evaluated the time-lag and duration models trained by the conventional part-isolated training and padding methods.
        
        Table~\ref{tab:padding} shows the results.
        Compared with Baseline, the prediction errors of time-lag and duration are superior with the proposed two padding methods. This result shows the effectiveness of taking into account the score-level interaction.
        Also the time-aligned padding method achieved the lowest prediction error, demonstrating the effectiveness of taking into account the synchrony across voice parts at the score level.

        \begin{table}[t]
    \centering
    \caption{Root mean squared errors of time-lag and duration. \textbf{Bold} means the best score}
    \label{tab:padding}
    \vspace{3mm}
    \footnotesize
    \begin{tabular}{c|c|c}\toprule
        Padding     & Time-lag [frame] $(\downarrow)$  & Duration [frame] $(\downarrow)$ \\ 
        \midrule
        Baseline  & $5.92$            & $23.08$ \\
        Post-padded & $5.89$            & $23.04$ \\ 
        Time-aligned& $\textbf{5.56}$            & $\textbf{22.86}$ \\ \bottomrule
    \end{tabular}
    \vspace{-3mm}
\end{table}

    \subsection{Evaluation of part-wise singing voices} \label{subsec:MOS_naturalness}
        Next, we conducted objective evaluation on synthesized singing voices of each part, using the following one baseline and four proposed methods.
        
        \smallskip
        \textbf{Baseline}: It uses the conventional voice-part-independent training method used in Section~\ref{subsec:音響モデルの比較}. That is, the interactions between singers are not explicitly utilized.
            
        \textbf{MT}: It uses our multi-track architecture (shown in Fig.~\ref{fig:architecture}) without the $F_0$ and power difference losses.
        Only the score-level interaction is explicitly modeled.
        
        \textbf{MT+F0diff}: It trains the MT model with the $F_0$ difference loss.
        
        \textbf{MT+Powdiff}: It trains the MT model with the power difference loss.
        
        \textbf{MT+F0diff+Powdiff}: It trains the MT model with the $F_0$ and power difference losses.
        It explicitly utilizes the interactions at the levels of score and acoustic feature.
        $\weight{F0diff}, \weight{Powdiff}$ were set to $1.0$.
        \smallskip
        
        We have not compared our methods with SVS methods other than Baseline, since the main focus of this study is to verify whether interaction mechanisms improves the unity of the vocal ensemble.
        

        \begin{table}[t]
    \centering
    \caption{Objective results of part-wise singing voices. \textbf{Bold} and \underline{underline} mean the best and second-best scores, respectively}
    \label{table:Objective}
    \vspace{3mm}
    \footnotesize
    \begin{tabular}{c|p{0.11\linewidth}|p{0.11\linewidth}|p{0.11\linewidth}|p{0.11\linewidth}}\toprule
        Method                  & MCD $(\downarrow)$   & LF0-RMSE $(\downarrow)$    & LF0diff-RMSE $(\downarrow)$   & Powdiff-RMSE $(\downarrow)$ \\ 
        \midrule
        Baseline                & $\mathbf{10.57}$     & $0.117$                   & $0.0443$                      & $1.457$ \\ 
        MT                      & $10.82$              & $\mathbf{0.100}$          & $0.0402$                      & $1.364$ \\ 
        MT$+$F0diff             & $10.88$              & $0.108$                   & $\mathbf{0.0389}$             & $1.317$\\ 
        MT$+$Powdiff            & $10.92$              & $\underline{0.104}$       & $0.0402$                      & $\mathbf{1.234}$\\ 
        MT$+$F0diff$+$Powdiff   & $\underline{10.67}$  & $0.106$                   & $\underline{0.0396}$          & $\underline{1.272}$ \\ \bottomrule
    \end{tabular}
 \end{table}


        Table~\ref{table:Objective} shows the results for four evaluation metrics: mel-cepstral distortion (MCD), and root mean square errors (RMSEs) of log $F_0$ (LF0-RMSE), log $F_0$ difference (LF0diff-RMSE), and power difference (Powdiff-RMSE). 
        Baseline had the lowest MCD but the highest LF0-RMSE, suggesting that it prioritizes phonetic features.
        Compared with Baseline, the three proposed methods provided better results for LF0-RMSE, LF0diff-RMSE, and Powediff-RMSE.
        The distinct feature between Baseline and the proposed methods is the use of the score features of another voice part.
        Since the note pitches of the voice parts are dependent on each other, their information should lead to the performance improvements.
        MT+F0diff and MT+Powdiff had lower LF0diff-RMSE and Powdiff-RMSE than MT, respectively.
        Furthermore, MT+F0diff+Powdiff achieved higher performances on average.
        This result demonstrates the effectiveness of the propsoed two losses, i.e., the importance of explicitly introducing the coherence in pitch and power across voice parts.
        

        \begin{figure}[t]
            \centering
            \includegraphics[width=0.95\linewidth]{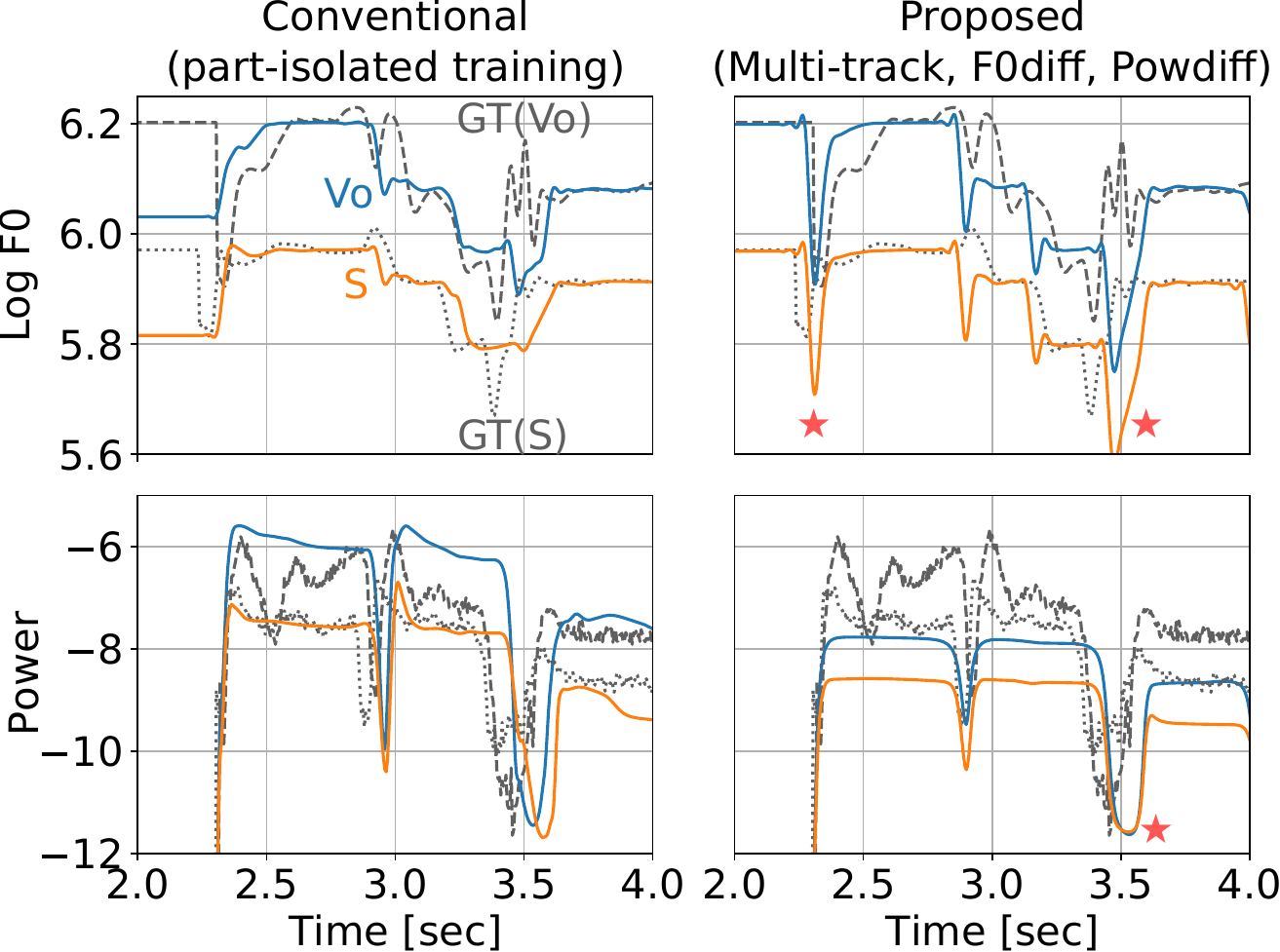}
            \vspace{-5pt}
            \caption{Ground-truth (GT) and predicted acoustic features by Baseline and proposed MT+F0diff+Powdiff. Vo and S denote lead vocal and soprano voice parts, respectively.
            Predicted features of proposed method synchronously change across voice parts, particularly at times marked with red stars.
            }
            \label{fig:f0-pow_example}
        \end{figure}

        Figure~\ref{fig:f0-pow_example} shows examples of log $F_0$ and power sequences generated by Baseline and proposed MT+F0diff+Powdiff.
        Since Baseline does not take into account the interaction between singers, the generated log $F_0$ and power sequences changed individually. For example, the two voice parts have different peak and dip in log $F_0$ at around 3.5 s in the left upper panel of Fig.~\ref{fig:f0-pow_example}.
        The acoustic features of the proposed method apparently changes synchronously across the voice parts.

    \subsection{Evaluation of the unity of vocal ensemble} 
        Finally, we conducted a five-point MOS test on the unity to evaluate the synthesized ensemble singing voices.
        For the MOS test, we adopted \textit{oneness} and \textit{voice harmony}, two of the six criteria for vocal ensemble mentioned in Section~\ref{Sec:関連研究}.
        We believe that this criterion is the easiest among the six criteria to evaluate, even for non-musical experts.
        60 listeners participated, and each listener assessed 20 phrases of singing voices.
        At the begging of the MOS test, the listeners listened to several dummy singing voices (including both natural and synthesized voices) to establish the rating scale.

        Table~\ref{table:MOS_matomari} shows the results. First, it can be seen that MT improves upon Baseline ($p = 0.22$ for oneness, $p = 0.34$ for voice harmony). The F0 difference loss showed favorable results in the LF0diff-RMSE presented in Table~\ref{table:Objective}, but it was found to worsen the MOS values.
        This is presumed to be due to the deterioration in LF0-RMSE shown in Table~\ref{table:Objective}. By contrast, the power difference loss improves MOS values, and MT+Powdiff is found to bring a significantly higher oneness ($p = 0.03$ for oneness, $p = 0.16$ for voice harmony) compared with Baseline.
        These results demonstrate the effectiveness of the proposed network architecture and the proposed loss function for power difference.

        \begin{table}[t]
    \centering
    \caption{MOS on unity of vocal ensemble. \textbf{Bold} indicates the best score and significantly better than baseline with $p < 0.05$.}
    \label{table:MOS_matomari}
    \vspace{3mm}
    \footnotesize
    \begin{tabular}{c|c|c}\toprule
        Method                  & \multicolumn{2}{|c}{MOS $\pm 95\%$ conf. $(\uparrow)$} \\ 
        & oneness & voice harmony\\ 
        \midrule
        Baseline                & $3.10\pm0.14$ & $3.25 \pm 0.14$\\ 
        MT                      & $3.22\pm0.14$ & $3.34 \pm 0.14$ \\ 
        MT$+$F0diff             & $2.92\pm0.14$ & $3.15 \pm 0.15$\\ 
        MT$+$Powdiff            & $\textbf{3.31}\pm0.13$ & $\textbf{3.38} \pm 0.14$\\ 
        MT$+$F0diff$+$Powdiff   & $2.89\pm0.14$ & $3.18 \pm 0.14$\\ \bottomrule
    \end{tabular}
    \vspace{-3mm}
 \end{table}

    \section{Conclusion}
        In this paper, we proposed a DNN-based ensemble singing voice synthesis method by explicitly modeling interaction between singers at the level of score and acoustic features.
        To induce the score-level interaction, the proposed network refers to the score features of all voice parts in addition to those of the target voice part.
        To induce the acoustic-feature-level interaction, we introduced loss functions of $F_0$ and power differences across voice parts.
        For training, we also proposed the time-aligned padding method for the score features, which captures the synchrony in note onset across voice parts.
        Experimental results showed that the proposed time-aligned padding method improves the prediction accuracy of time-lag and duration.
        They also demonstrated the effectiveness of the proposed method and the importance of explicitly modeling the score- and acoustic-feature-level interactions.
        Future work includes further improvements of unity of the synthesized ensemble singing voices and elucidation of how humans perceive unity within ensemble voices.

    \section{Acknowledgments}
        This work was supported by JST FOREST JPMJFR226V, and JSPS KAKENHI Grant Numbers JP23K28108 and JP23K18474.




\bibliographystyle{IEEEbib_original}
\bibliography{strings,refs}

\end{document}